# Models with time-dependent parameters using transform methods: application to Heston's model

A. Elices[1]

*Abstract*— **This paper presents a methodology to introduce time-dependent parameters for a wide family of models preserving their analytic tractability. This family includes hybrid models with stochastic volatility, stochastic interest-rates, jumps and their non-hybrid counterparts. The methodology is applied to Heston's model. A bootstrapping algorithm is presented for calibration. A case study works out the calibration of the time-dependent parameters to the volatility surface of the Eurostoxx 50 index. The methodology is also applied to the analytic valuation of forward start vanilla options driven by Heston´s model. This result is used to explore the forward skew of the case study.**

## I. INTRODUCTION

Long term hybrids are drawing considerable attention to both sales people and traders in the past year as the uncertainty of the market fosters clients to hedge their risks through longer terms. Most of these complex exotic derivatives are typically valued using Monte-Carlo methods. However, calibration of model parameters to market involves many evaluations of vanilla products. Therefore, an analytic or at least a quick evaluation method for vanilla options is crucial. One of the major drawbacks of analytic models is that they depend on just a few parameters which do not provide enough degrees of freedom to fit the market at several maturities. The whole motivation of this paper is to provide more degrees of freedom by introducing piecewise constant time-dependent parameters.

This problem is not new and several authors have already given solutions for specific models. The main contribution of this paper is a methodology to extend not only a specific model, but a wide family of them so that time-dependent parameters can be introduced preserving analytic tractability. This family includes models with stochastic volatility (e.g. [Heston, 1993]), jumps (e.g. [Merton, 1976]) and hybrid models with stochastic volatility and stochastic interest rates correlated with the underlying (e.g. [Bakshi, 1997] and [Scott, 1997]). The methodology is based on characteristic function methods which describe the probability distributions of the stochastic processes in terms of the characteristic function. This general methodology is illustrated by applying it to Heston's model for valuation of both spot and forward start vanilla options. This same problem has already been addressed by [Mikhailov et al, 2005] from a completely different perspective using partial differential equations. In the latter paper, the solution to Heston´s partial differential equation is obtained for successive periods with different sets of parameters. The solution of the last period is used as initial condition for the preceding one. The solution of this period is applied to the preceding one until the first period is reached. [Sepp, 2008] also addresses the problem of pricing Vix futures and options on the Vix index through a model with time-dependent parameters. The Vix index measures the implied volatility of S&P 500 with a maturity of 30 days. The model assumes stochastic volatility with positive exponentially distributed jumps to fit an upward sloping skew. One of the key results of this paper is the derivation of the characteristic function with time-dependent parameters throgh recursion. Again, this derivation is based on cumbersome partial differencial equations and could be easily solved using the methodology proposed in this paper.

Another very interesting related work may be found in [Piterbarg, 2005] and [Piterbarg, 2006]. These papers derive approximate formulas to imply time-dependent parameters in between maturities from average parameters which fit market prices at each maturity. This allows an independent calibration of average parameters to fit market prices at each maturity (this data may already be available in trading desks). From these average parameters, a time-dependent model is implied. This model replicates the distribution and market prices at each maturity. The specific model used to describe the underlying process is the displaced diffusion stochastic volatility model in [Andersen et al, 2002]. A local volatility function controls the slope of the implied volatility smile allowing independent Brownian motions for the stochastic volatility and the underlying process.

[Britten-Jones et al, 2000] prove that for all continuous processes the expected value of realized variance up to a given maturity is defined by vanilla option prices for that maturity with respect to a continuum of strikes. This means that all possible models which calibrate vanilla option prices must have the same expected value of realized variance. For instance, Dupire's local volatility model fits the expected value of realized variance by a deterministic function (the local volatility). It is important to have in mind that exotic pricing has to be taken with care and prices given by any model should be considered in its right context. [Schoutens et al, 2004] or [Britten-Jones et al, 2000] give some examples of how a variety of different models fitting the smile option prices give considerably different prices depending on the hypothesis of the underlying process. Section V works out two different calibration sets which fit market prices. They will be compared in section VII in terms of the forward skew and the price differences will be explained. Most traders and practitioners like to compare the prices of different models and trade and hedge with the model which gives the closest price to market (or what they think that should be the correct market price).

---

[1] Senior Quant member of the model validation group at Grupo Santander. Email: aelices@gruposantander.com. Address: Área Metodología, Ed. Encinar P2, Avda Cantabria s/n, Ciudad Grupo Santander, 28660 Boadilla del Monte, Spain.

The paper is organized in five sections. Section II (the main contribution of the paper) presents a general methodology to derive characteristic functions for a time horizon where the parameters of the underlying process may change. This characteristic function is expressed in terms of the characteristic functions of each sub-period where parameters change. Section III applies the methodology to the well-known model by Heston and section IV proposes a bootstrapping calibration algorithm. Section V presents a case study which works out two different calibration sets of a Heston model with time-dependent parameters to the volatility surface of the Eurostoxx 50 index. Section VI applies again the methodology to derive a semi-analytic formula for the valuation of forward start vanilla options driven by Heston´s model. Section VII uses this formula to explore the forward skew of both calibration sets provided by section V and explaing why they give different results. The paper ends with some conclusions. Appendix A derives a more general version of Heston's characteristic function so that the new methodology can be applied.

## II. CHARACTERISTIC FUNCTIONS OF MODELS WITH TIME DEPENDENT PARAMETERS

Consider the characteristic function (1) of the distribution of a Markov $N$-dimensional process $\mathbf{x}(t) = (x_1(t), \cdots, x_N(t))$. From a mathematical point of view, the characteristic function is the Fourier Transform of the density function.

$$\varphi_{uv}(\mathbf{X}/\mathbf{x}_u) = \mathbf{E}\left(e^{i\mathbf{X}\cdot\mathbf{x}_v}\right) = \int_{\mathbf{R}^N} e^{i\mathbf{X}\cdot\mathbf{x}_v} f_{uv}(\mathbf{X}/\mathbf{x}_u) d\mathbf{x} \qquad (1)$$

The notation $\varphi_{uv}(\mathbf{X}/\mathbf{x}_u)$ and $f_{uv}(\mathbf{X}/\mathbf{x}_u)$ refers to the characteristic function and the density function of the joint distribution of the process $\mathbf{x}_v = \mathbf{x}(t_v)$ at time $t_v$, conditioned by its initial value $\mathbf{x}_u = \mathbf{x}(t_u)$ at $t_u$. $\varphi_{uv}(\mathbf{X}/\mathbf{x}_u)$ is a function of the vector $\mathbf{X} = (X_1, \cdots, X_N)$ and the notation $\mathbf{X} \cdot \mathbf{x}_v = \sum X_k x_k(t_v)$ refers to the inner product of vectors $\mathbf{X}$ and $\mathbf{x}_v$.

Consider now the family of exponential characteristic functions of the form (2) with exponents linear in the stochastic processes $\mathbf{x}_u$ at time $t_u$. The vector function $\mathbf{D}_{uv}(\mathbf{X}) = (D_{uv,1}(\mathbf{X}), \cdots, D_{uv,N}(\mathbf{X}))$ and the function $C_{uv}(\mathbf{X})$ depend not only on $\mathbf{X}$, but on the parameters of the particular model under consideration in the period from $t_u$ to $t_v$. This parameter dependence is dropped to simplify notation.

$$\varphi_{uv}(\mathbf{X}/\mathbf{x}_u) = \exp\left(C_{uv}(\mathbf{X}) + \mathbf{D}_{uv}(\mathbf{X}) \cdot \mathbf{x}_u\right) \qquad (2)$$

A wide range of models belong this class. Examples are provided below. Equation (3) presents the form of the characteristic function of Merton's lognormal jump diffusion model [Merton, 1976]. The variable $g_u$ is the sum of all jumps (which happen randomly following a Poisson process) up to time $t_u$.

$$\varphi_{uv}(G/g_u) = \exp\left(C_{uv}(G) + iG g_u\right) \qquad (3)$$

The form of the characteristic function of the fixed income short rate model by Cox, Ingersoll and Ross [Cox et al, 1985a] is given by equation (4), where $r_u$ is the short rate at time $t_u$.

$$\varphi_{uv}(R/r_u) = \exp\left(C_{uv}(R) + iD_{uv}(R)r_u\right) \qquad (4)$$

Similarly, the form of the characteristic function of Heston's lognormal model with stochastic volatility [Heston, 1993] is presented by equation (5), where $x_u$ is the logarithm of the underlying stock and $v_u$ is the variance (both at time $t_u$).

$$\varphi_{uv}(X,V/x_u,v_u) = \exp\left(C_{uv}(X,V) + D_{uv}(X,V)v_u + iXx_u\right) \qquad (5)$$

An example of a hybrid model combining the latter four altogether such as [Bakshi, 1997] or [Scott, 1997] is given by equation (6), where $C$ and $D$ depend on the four variables $X, V, R$ and $G$.

$$\varphi_{uv}(X,V,R,G/x_u,v_u,r_u,g_u) = e^{C_{uv}+D_{uv,2}r_u+D_{uv,1}v_u+Xx_u+Gg_u} \qquad (6)$$

All these models allow for quick semi-analytic formulas to price vanilla options by using transform methods that invert the characteristic function. The main drawback of these models is that they depend on a few parameters which do not provide enough degrees of freedom to calibrate the market. The goal of this section is to derive the characteristic function of a process that may have time-dependent parameters.

As vanilla options used for calibration are usually European, all the information of a Markov process with independent increments up to an instant is given by the joint probability distribution of the stochastic variables which describe it at that instant (marginal distributions can always be calculated from the joint distribution). In addition, all the information necessary to continue the evolution of this process from an instant $t_u$ to a later one $t_v$, is the joint distribution at $t_u$ and the evolution law from $t_u$ to $t_v$. Fig. 1 represents graphically this information: from 0 to $t_u$ the process is described by the characteristic function $\varphi_{0u}(\mathbf{X}/\mathbf{x}_0)$ conditioned by $\mathbf{x}_0$ and from $t_u$ to $t_v$, the process is described by $\varphi_{uv}(\mathbf{X}/\mathbf{x}_u)$ conditioned by $\mathbf{x}_u$. Note that both characteristic functions may represent the underlying evolution with different parameters. In this context, the goal is to obtain the characteristic function $\varphi_{0v}(\mathbf{X}/\mathbf{x}_0)$ of the joint distribution at $t_v$ given $\mathbf{x}_0$ in terms of $\varphi_{0u}(\mathbf{X}/\mathbf{x}_0)$ and $\varphi_{uv}(\mathbf{X}/\mathbf{x}_u)$.

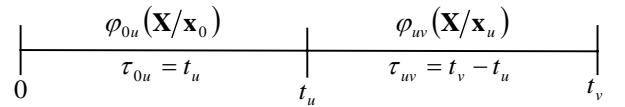

Fig. 1: Graphical representation of a Markov process with independent increments in two consecutive periods.

Equation (7) shows the definition of the characteristic function under search.

$$\varphi_{0v}(\mathbf{X}/\mathbf{x}_0) = \int_{\mathbf{R}^N} d\mathbf{x}_v e^{i\mathbf{X}\cdot\mathbf{x}_v} f_{0v}(\mathbf{x}_v/\mathbf{x}_0) \qquad (7)$$

As non-overlapping intervals are independent, the density from 0 to $t_v$ is the product of the densities from 0 to $t_u$ and from $t_u$ to $t_v$, summed over all intermediate paths $\mathbf{x}_u$ as shown in equation (8).

$$f_{0v}(\mathbf{x}_v/\mathbf{x}_0) = \int_{\mathbf{R}^N} d\mathbf{x}_u f_{0u}(\mathbf{x}_u/\mathbf{x}_0) f_{uv}(\mathbf{x}_v/\mathbf{x}_u) \qquad (8)$$

Substituting (8) in (7) and exchanging the sum order (summing over $x_v$ first) yields equation (9):

$$\varphi_{0v}(\mathbf{X}/\mathbf{x}_0) = \int_{\mathbf{R}^N} d\mathbf{x}_u f_{0u}(\mathbf{x}_u/\mathbf{x}_0) \int_{\mathbf{R}^N} d\mathbf{x}_v e^{i\mathbf{X}\cdot\mathbf{x}_v} f_{uv}(\mathbf{x}_v/\mathbf{x}_u) \quad (9)$$

As the second integral of (9) is the definition of $\varphi_{uv}(\mathbf{X}/\mathbf{x}_u)$, equation (9) becomes (10):

$$\varphi_{0v}(\mathbf{X}/\mathbf{x}_0) = \int_{\mathbf{R}^N} d\mathbf{x}_u f_{0u}(\mathbf{x}_u/\mathbf{x}_0) \varphi_{uv}(\mathbf{X}/\mathbf{x}_u) \quad (10)$$

Substituting equation (2) in equation (10), applying the definition of $\varphi_{0u}(\mathbf{X}/\mathbf{x}_0)$ in (12) and substituting $\varphi_{0u}(\mathbf{X}/\mathbf{x}_0)$ according to equation (2) in (13) yields:

$$\varphi_{0v}(\mathbf{X}/\mathbf{x}_0) = \int_{\mathbf{R}^N} d\mathbf{x}_u f_{0u}(\mathbf{x}_u/\mathbf{x}_0) \exp(C_{uv}(\mathbf{X}) + \mathbf{D}_{uv}(\mathbf{X})\cdot \mathbf{x}_u) \quad (11)$$

$$= \exp(C_{uv}(\mathbf{X})) \int_{\mathbf{R}^N} d\mathbf{x}_u f_{0u}(\mathbf{x}_u/\mathbf{x}_0) \exp(i(i^{-1}\mathbf{D}_{uv}(\mathbf{X}))\cdot \mathbf{x}_u) \quad (12)$$

$$= \exp(C_{uv}(\mathbf{X})) \varphi_{0u}(i^{-1}\mathbf{D}_{uv}(\mathbf{X})/\mathbf{x}_0) \quad (13)$$

$$= \exp(C_{uv}(\mathbf{X}) + C_{0u}(i^{-1}\mathbf{D}_{uv}(\mathbf{X})) + \mathbf{D}_{0u}(i^{-1}\mathbf{D}_{uv}(\mathbf{X}))\cdot \mathbf{x}_0) \quad (14)$$

Applying equation (2) to the interval 0 to $t_v$ yields (15):

$$\varphi_{0v}(\mathbf{X}/\mathbf{x}_0) = \exp(C_{0v}(\mathbf{X}) + \mathbf{D}_{0v}(\mathbf{X})\cdot \mathbf{x}_0) \quad (15)$$

Identifying terms between equations (14) and (15) yields (16): the expression of $\varphi_{0v}(\mathbf{X}/\mathbf{x}_0)$ in terms of $\varphi_{0u}(\mathbf{X}/\mathbf{x}_0)$ and $\varphi_{uv}(\mathbf{X}/\mathbf{x}_u)$.

$$\begin{cases} C_{0v}(\mathbf{X}) = C_{uv}(\mathbf{X}) + C_{0u}(i^{-1}\mathbf{D}_{uv}(\mathbf{X})) \\ \mathbf{D}_{0v}(\mathbf{X}) = \mathbf{D}_{0u}(i^{-1}\mathbf{D}_{uv}(\mathbf{X})) \end{cases} \quad (16)$$

Consider now Fig. 2 with a series of periods in which the parameters of the process are different.

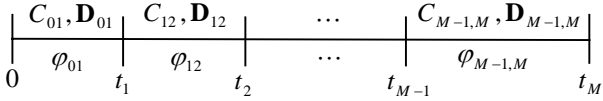

Fig. 2: Example of a five period process.

The characteristic function $\varphi_{0M}$ at a given maturity $t_M$ can be obtained recursively applying equation (16) to $\varphi_{M-1,M}$, given by equation (2), and $\varphi_{0,M-1}$. $\varphi_{0,M-1}$ is obtained applying again equation (16) to $\varphi_{M-2,M-1}$, given by equation (2), and $\varphi_{0,M-2}$. This process continues until $\varphi_{02}$ is obtained applying equation (16) to $\varphi_{01}$ and $\varphi_{12}$ where both are calculated using equation (2).

The marginal distribution of the $h$th stochastic variable can be calculated from the definition of characteristic function by setting to zero all the $X_k$ except for $X_h$ as indicated in equation (17).

$$\varphi_{0v}(X_h/\mathbf{x}_0) = \varphi_{0v}((X_1,\cdots,X_N)/\mathbf{x}_0)\big|_{X_k=0 \; k\neq h} \quad (17)$$

III. APPLICATION TO HESTON'S MODEL

Equation (18) presents Heston's underlying process $S_t$,

$$\begin{cases} dS_t = \mu S_t dt + \sqrt{v_t} S_t dW_t \\ dv_t = \kappa(\theta - v_t)dt + \sigma\sqrt{v_t} dY_t \end{cases} \quad d\langle W_t, Y_t\rangle = \rho dt \quad (18)$$

where $v_t$ is the stochastic variance. This variance follows an Ornstein-Uhlenbeck process where $\kappa$ is the mean reversion rate, $\theta$ is the long term variance and $\sigma$ is the volatility of the variance process. There is a correlation $\rho$ between the Brownian motions which drive the underlying process and the variance. The parameter $\mu$ is the risk-neutral drift[2].

$$C = P(0,T)\mathbf{E}((S_T - K)^+) \quad (19)$$

Equation (19) shows the price of a call option where $K$ is the strike price, $P(0,T)$ is the discount factor from expiry $T$ to present time, the expectation is calculated with the information of present time and the measure of the expectation is the same used to calculate the risk neutral drift $\mu$ of equation (18).

$$C = P(0,T)\mathbf{E}(e^{x_T}\mathbf{1}_{\{x_T>\ln K\}}) - P(0,T)K\mathbf{E}(\mathbf{1}_{\{x_T>\ln K\}}) \quad (20)$$

Equation (20) breaks the price into two expectations and expresses the payoff in terms of $x_t = \ln S_t$. In order to calculate these expectations, it is necessary to know the marginal distribution of $x_T$ given $x_0$ and $v_0$. Although the density function of this distribution is not analytic, Heston derived an analytic expression for the characteristic function [Heston, 1993]. In order to apply the methodology developed in section II, the characteristic function of the joint distribution of $x_T$ and $v_T$ given $x_0$ and $v_0$ (and not the marginal) must be calculated. Appendix A presents the details of this calculation. The final result for the period from $t_u$ to $t_v$ is given by equation (21),

$$\varphi_{uv}(\mathbf{X}/\mathbf{x}_u) = e^{C_{uv}(\mathbf{X}) + D_{uv,2}(\mathbf{X})v(t_u) + D_{uv,1}(\mathbf{X})x(t_u)} \quad (21)$$

where $\mathbf{X} = (X,V)$, $\mathbf{x}_u = (x(t_u), v(t_u))$. $C_{uv}(\mathbf{X})$ is given by equation (76) (with $C^0 = 0$ and $D^0 = iV$). $D_{uv,2}(\mathbf{X})$ is given by equation (69), $D_{uv,1}(\mathbf{X}) = iX$ and $\tau = t_v - t_u$. The variable $X$ of the characteristic function corresponds to the logarithm of the underlying stock and $V$ corresponds to the variance process.

$$\begin{cases} C_{0v}(X,V) = C_{uv}(X,V) + C_{0u}(X,i^{-1}D_{uv,2}(X,V)) \\ D_{0v,2}(X,V) = D_{0u,2}(X,i^{-1}D_{uv,2}(X,V)) \\ D_{0v,1}(X,V) = iX \end{cases} \quad (22)$$

If time dependent parameters across several periods are considered, equation (16) becomes (22) and the same procedure of section II can be applied to get the joint characteristic function (23) from 0 to any time.

$$\varphi_{0v}(X,V/x_0,v_0) = e^{C_{0v}(X,V) + D_{0v,2}(X,V)v_0 + D_{0v,1}(X,V)x_0} \quad (23)$$

The marginal characteristic function of $x_T$ is obtained by simply evaluating the joint characteristic function at $V = 0$ according to equation (24).

$$\varphi_{0T}(X/x_0,v_0) = \varphi_{0T}(X,0/x_0,v_0) = \mathbf{E}(e^{iXx_T}) \quad (24)$$

Equation (25) presents the inversion formula to calculate the probability $P$ from a distribution defined by its characteristic function $\varphi$. This formula is found in [Kendal, 1987] and [Shephard, 1991a]. A discussion on Fourier inversion formulas is also available in [Feller, 1971].

---

[2] By risk-neutral drift it is meant the drift that forces the process $S_t/N_t$ to be a martingale under the chosen numeraire $N_t$. Throughout this paper the numeraire $N_t$ will be the bank account $B_t$.=exp(r×t) Under these assumptions, $\mu$=r-q where r is the risk free rate and q is the continuous dividend yield of the stock.

$$P(x>a)=\frac{1}{2}+\frac{1}{2\pi}\int_0^\infty \frac{1}{iX}\left(\frac{\varphi(X)}{e^{iXa}}-\frac{\varphi(-X)}{e^{-iXa}}\right)dX \quad (25)$$

When $\varphi(-X)$ and $\varphi(X)$ are complex conjugates equation (25) can be reduced to (26), where **Re**(.) stands for the real part (this is the final result given in [Heston, 1993]). It can be verified that this condition is satisfied for both equation (21) (flat model) and (22) (model with time dependent parameters).

$$P(x>a)=\frac{1}{2}+\frac{1}{\pi}\int_0^\infty \mathbf{Re}\left(\frac{\varphi(X)e^{-iXa}}{iX}\right)dX \quad (26)$$

The second expectation in (20) is calculated using the inversion formula (25) with $\varphi(X)=\varphi_{0T}(X/\mathbf{x}_0)$ and $a=\ln K$. The first expectation in (20) can be obtained similarly using the characteristic function $\tilde{\varphi}_{0T}(X/\mathbf{x}_0)$ of equation (27).

$$\tilde{\varphi}_{0T}(X/\mathbf{x}_0)=\frac{\varphi_{0T}(X-i/\mathbf{x}_0)}{\varphi_{0T}(-i/\mathbf{x}_0)}=\frac{\mathbf{E}\left(e^{i(X-i)x_T}\right)}{\mathbf{E}\left(e^{x_T}\right)}=\frac{\mathbf{E}\left(e^{iXx_T+x_T}\right)}{\mathbf{E}\left(e^{x_T}\right)} \quad (27)$$

To apply the inversion formula (25), it is necessary that $\tilde{\varphi}_{0T}(0/\mathbf{x}_0)$ equals one, so that $\tilde{\varphi}_{0T}(0/\mathbf{x}_0)$ be a characteristic function of a fictitious density $\tilde{f}_{0T}(0/\mathbf{x}_0)$ that sums one over the whole domain. Therefore, $\varphi_{0T}(X-i/\mathbf{x}_0)$ is normalized by the constant $\varphi_{0T}(-i/\mathbf{x}_0)$ (the forward price of the underlying).

$$\mathbf{E}\left(e^{x_T}\mathbf{1}_{\{x_T>\ln K\}}\right)=\mathbf{E}(S_T)\tilde{P}(x_T>\ln K) \quad (28)$$

The probability $\tilde{P}$ given by formula (25) with $\varphi(X)=\tilde{\varphi}_{0T}(X/\mathbf{x}_0)$ and $a=\ln K$ will yield the desired expectation normalized by the forward price. Therefore the expectation will be given by equation (28) and the final call price by (29):

$$C=P(0,T)S_0 e^{\mu T}\tilde{P}(x_T>\ln K)-P(0,T)KP(x_T>\ln K) \quad (29)$$

Fast and accurate methods to implement the inversion formula (25) can be found in [Shephard, 1991b] and [Davies, 1973]. A well-known algorithm to avoid the singularity at zero of (25) and apply the Fast Fourier Transform algorithm controlling the precision is described in [Lee, 2005] and [Carr et al, 1998].

## IV. CALIBRATION

The marginal density function of the underlying at a given maturity is completely determined by a continuum of vanilla prices dependent on the strike. A good approximation of this distribution can be obtained from the interpolation of the implied volatilities of vanilla calls and puts with respect to the strike. As more exotic path-dependent options are not as liquid as vanillas and they are usually traded over the counter, the market provides quite limited information about the evolution of the underlying process in between maturities. In addition, introducing path-dependent products in the calibration is quite challenging because analytic solutions are usually not available. Therefore, any model calibrated to market should at least reproduce the marginal distribution of the underlying at the maturities for which vanilla products are quoted.

The most immediate and probably the simplest and quickest solution to calibrate a model with time-dependent parameters is a bootstrapping algorithm. The periods where the parameters change are given by the periods in between the maturities of the vanilla products used for calibration. Each period is calibrated independently starting from the first to the last solving a minimisation problem with the objective function of equation (30). The weights $w_i$ are chosen to give the highest priority to the at the money (ATM) options. These weights will decrease as the moneyness of the option gets apart from ATM.

$$FO=\sum_{i=1}^M \frac{w_i}{\sum w_j}\left(price_i^{model}-price_i^{market}\right)^2 \quad (30)$$

When the parameters of the first period are calibrated to fit vanilla prices expiring on the first maturity, they are fixed. The parameters of the second period are then calibrated to fit vanilla prices expiring on the second maturity leaving the parameters of the first period fixed. Now, the parameters of the first two periods are fixed and the parameters of the third period are calibrated to fit vanilla prices expiring on the third maturity. This process continues until the last period. The advantage of a bootstrapping algorithm is that each period involves an optimisation with only the parameters of that period.

## V. CASE STUDY: CALIBRATION OF THE EUROSTOXX 50 INDEX

The whole methodology proposed in this paper is applied to Heston´s model for the calibration of the Eurostoxx 50 index. The spot price is 3868.64€ and the volatility surface is given by Table 1. The leftmost column shows the moneyness of the options with respect to the spot price (the strikes are the moneyness times the spot).

TABLE 1: EUROSTOXX 50 VOLATILITY SURFACE.

| K \ Mat | 1m | 3m | 6m | 9m | 1y | 2y | 3y | 4y | 5y | 10y |
|---|---|---|---|---|---|---|---|---|---|---|
| 0.85 | 23.0 | 18.7 | 18.5 | 18.6 | 19.1 | 19.7 | 20.6 | 21.5 | 22.2 | 25.8 |
| 0.90 | 18.9 | 16.7 | 17.0 | 17.2 | 17.8 | 18.8 | 19.8 | 20.8 | 21.5 | 25.3 |
| 0.95 | 15.2 | 14.7 | 15.5 | 16.0 | 16.6 | 17.8 | 19.0 | 20.0 | 20.8 | 24.7 |
| 1.00 | 12.2 | 13.2 | 14.1 | 14.8 | 15.5 | 16.9 | 18.2 | 19.3 | 20.2 | 24.2 |
| 1.05 | 11.6 | 12.3 | 13.1 | 13.9 | 14.4 | 16.1 | 17.5 | 18.7 | 19.5 | 23.7 |
| 1.10 | 13.3 | 12.3 | 12.6 | 13.2 | 13.7 | 15.4 | 16.9 | 18.1 | 19.0 | 23.2 |
| 1.15 | 15.6 | 12.9 | 12.4 | 12.7 | 13.2 | 14.8 | 16.3 | 17.5 | 18.5 | 22.7 |

To avoid problems with discrete dividend payments, it is more convenient to calibrate the model of the forward price of the underlying $F_t^P$ delivered on the last maturity date $P$, rather than the underlying spot $S_t$. When pricing exotics by Monte Carlo, the evolution of the forward is simulated and the spot price is recovered from the forward at each time using equation (31), where *NPV* is the net present value of all discrete dividends from $t$ to the delivery date $P$ of the forward.

$$F_t^P=\frac{S_t-NPV(dividends_{t-P})}{P(t,P)} \quad (31)$$

The prices of the vanilla options on $S_t$ should also be replaced by equivalent vanilla options on $F_t^P$ according to equation (32). If both interest rates and dividends are deterministic, the right hand side of (32) follows by multiplying and dividing the left hand side by the constant $F_0^{T_i}/F_0^P$.

$$E\left((S_{T_i} - K)^+\right) = \frac{F_0^{T_i}}{F_0^P} E\left(\left(F_{T_i}^P - K \frac{F_0^P}{F_0^{T_i}}\right)^+\right) \quad (32)$$

This is equivalent to consider that the implied volatilities of options on the spot $S_t$ with strike $K$ are the same as the implied volatilities on options on the forward $F_t^P$ with an adjusted strike equal to $K F_0^P / F_0^{T_i}$ (note that these strikes change for each maturity). Table 2 shows the forward values $F_0^{T_i}$ valued at present time and delivered at each maturity $T_i$. From this table $F_0^P = 4107.9 \, €$

TABLE 2: UNDERLYING FORWARD AT EACH MATURITY.

| 1m | 3m | 6m | 9m | 1y | 2y | 3y | 4y | 5y | 10y |
|---|---|---|---|---|---|---|---|---|---|
| 3870.6 | 3874.4 | 3880.3 | 3886 | 3892 | 3915.3 | 3938.9 | 3962.6 | 3986.5 | 4107.9 |

The bootstrapping algorithm of section IV is implemented using the function "fminsearch" of the scientific package Matlab. The first calibration step searches for the parameters of the first period and the initial variance $v_0$. The initial variance is fixed at this first step for the rest of the calibrations. The weights of equation (30) are set to 100 for ATM options and 45, 35 and 5 as the moneyness gets apart the ATM. Call options are used for strikes greater than $F_0^P$ and put options for strikes below. It has been observed that very different sets of parameters can fit the same market prices. This fact is not surprising as for a given volatility of variance, a sufficiently big mean reversion can produce the same result as a low mean reversion with very low volatility of variance. Therefore, the parameter seach space is limited to a set of intervals defined by the user so that more sensible parameters get out of the calibration. Table 3 presents the two search spaces that are considered. The set of intervals on the left represents a constrained search (especially with respect to $\sigma$ and $\kappa$), whereas the set on the right represents an uncontrained search.

TABLE 3: SEARCH SPACE: CONSTRAINED (LEFT), UNCONSTRAINED (RIGHT).

|     | $v_0$ | $\theta$ | $\kappa$ | $\sigma$ | $\rho$ | $v_0$ | $\theta$ | $\kappa$ | $\sigma$ | $\rho$ |
|-----|---|---|---|---|---|---|---|---|---|---|
| max | 1 | 1 | 20 | 1.5 | 1 | 100 | 100 | 100 | 100 | 1 |
| min | 0 | 0 | 0 | 0 | -1 | 0 | 0 | 0 | 0 | -1 |

To avoid constrained optimisation, equation (33) shows a change of variable for which the constrained parameter $p$ is expressed in terms of an unconstrained $\tilde{p}$ for which the search is carried out. The constants $p_{\min}$ and $p_{\max}$ are the limits of the interval in which the parameter $p$ is confined when $\tilde{p}$ moves in the real line. The constant $m$ has been set to 100 to make the transition of the hyperbolic tangent from –1 to 1 less abrupt.

$$p = \frac{p_{\max} - p_{\min}}{2}\left(1 + p_{\min} + \tanh\left(\frac{\tilde{p}}{m}\right)\right) \quad (33)$$

TABLE 4: CALIBRATED HESTON PARAMETERS: CONSTRAINED (UP), UNCONSTRAINED (DOWN).

| P \ Mat | 1m | 3m | 6m | 9m | 1y | 2y | 3y | 4y | 5y | 10y |
|---|---|---|---|---|---|---|---|---|---|---|
| $\theta$ | 0.01 | 0.03 | 0.03 | 0.03 | 0.05 | 0.05 | 0.07 | 0.12 | 0.14 | 0.31 |
| $\kappa$ | 0.61 | 7.33 | 6.25 | 6.46 | 4.20 | 2.78 | 1.97 | 0.84 | 0.61 | 0.29 |
| $\sigma$ | 0.60 | 0.56 | 1.13 | 1.15 | 1.09 | 1.26 | 1.18 | 1.14 | 1.12 | 1.14 |
| $\rho$ | -0.42 | -0.46 | -0.59 | -0.63 | -0.90 | -0.67 | -0.75 | -0.77 | -0.79 | -0.84 |
| $\theta$ | 0.01 | 0.03 | 0.03 | 0.03 | 0.05 | 0.05 | 0.06 | 0.09 | 0.11 | 0.21 |
| $\kappa$ | 0.84 | 4.75 | 3.08 | 5.21 | 4.87 | 4.82 | 3.81 | 3.89 | 4.51 | 3.02 |
| $\sigma$ | 0.61 | 0.35 | 0.77 | 0.83 | 1.54 | 1.58 | 1.88 | 3.35 | 5.24 | 6.70 |
| $\rho$ | -0.42 | -0.57 | -0.56 | -0.68 | -0.77 | -0.78 | -0.80 | -0.85 | -0.88 | -0.92 |

Table 4, Fig. 3 and Fig. 4 present the calibrated parameters for the constrained $v_0 = 0.0174$ and the unconstrained $v_0 = 0.0175$ cases. The undiscounted vanilla option prices on $F_t^P$ with spot equal to $F_0^P$ and adjusted strike prices $K F_0^P / F_0^{T_i}$ for each moneyness and maturity are presented in Table 5. These prices are normalized by $F_0^P$, and expressed in basis points.

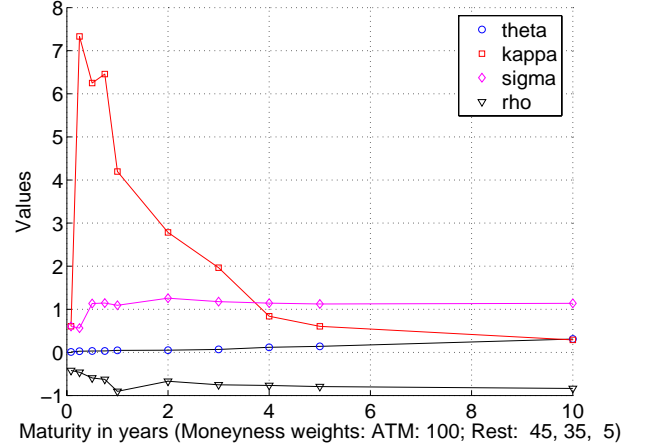

Fig. 3: Calibrated parameters with **constrained** search space.

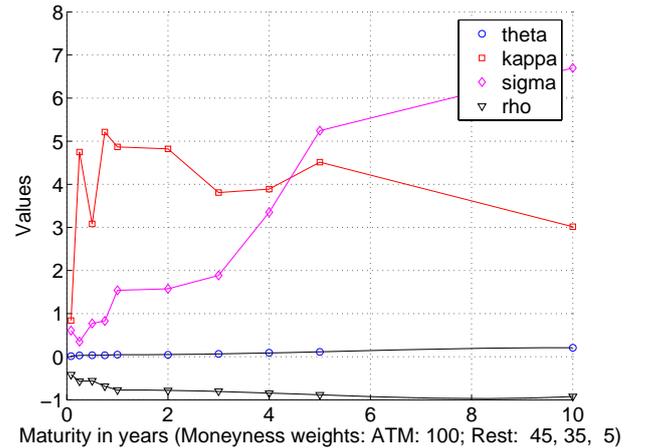

Fig. 4: Calibrated parameters with **unconstrained** search space.

Table 6 shows the calibration error (market minus Heston) for each option in basis points for the constrained (above) and the unconstrained (below) cases. Note that these errors are all below 4 basis points except for the most out of the money options at long maturities. Both calibrations seem reasonable.

The interpretation of the time evolution of the parameters in terms of market expectations is tricky. Both calibrations suggest that the market is pricing in increasing volatility (increasing $\theta$ from short term volatility levels around 11% to 45% for 10 year maturity) and increasing skew ($\rho$ progressively getting closer to –1). The unconstrained calibration suggests increasing uncertainty for the volatility (increasing $\sigma$) as the mean reversion is rather stable around 4. The constrained calibration forces the volatility of variance $\sigma$ to be rather stable (the maximum level is 1.5) but the mean reversion $\kappa$ progressively decreases indicating more long term uncertainty for the volatility. Therefore, from a qualitative point of view, both calibrations seem to agree that market is pricing in increasing volatility, increasing uncertainty for the volatility and increasing skew

(more probability for outcomes with lower underlying levels).

TABLE 5: VANILLA OPTION PRICES (BASIS POINTS).

| K \ Mat | 1m | 3m | 6m | 9m | 1y | 2y | 3y | 4y | 5y | 10y |
|---|---|---|---|---|---|---|---|---|---|---|
| 0.85 | 0.4 | 12.8 | 60.1 | 115 | 181 | 412 | 632 | 850 | 1045 | 1958 |
| 0.90 | 3.7 | 36.8 | 115 | 190 | 271 | 537 | 780 | 1011 | 1216 | 2166 |
| 0.95 | 25.3 | 100 | 213 | 308 | 401 | 694 | 953 | 1194 | 1405 | 2384 |
| 1.00 | 138 | 255 | 383 | 488 | 584 | 887 | 1155 | 1399 | 1614 | 2612 |
| 1.05 | 11.7 | 79.7 | 189 | 295 | 396 | 747 | 1074 | 1378 | 1653 | 2851 |
| 1.10 | 0.7 | 18.5 | 72.2 | 144 | 220 | 532 | 846 | 1145 | 1418 | 2742 |
| 1.15 | 0.0 | 4.3 | 25.5 | 63.7 | 111 | 362 | 652 | 938 | 1205 | 2532 |

TABLE 6: CALIBRATION ERROR (BASIS POINTS): CONSTRAINED (UP), UNCONTRAINED (DOWN).

| K \ Mat | 1m | 3m | 6m | 9m | 1y | 2y | 3y | 4y | 5y | 10y |
|---|---|---|---|---|---|---|---|---|---|---|
| 0.85 | 1 | 1 | -2 | 0 | 1 | -4 | 0 | -1 | -3 | -4 |
| 0.90 | 2 | 1 | -1 | -1 | 0 | 0 | 0 | 1 | 0 | -1 |
| 0.95 | -1 | -1 | 1 | 0 | 0 | 1 | 0 | 0 | 0 | -1 |
| 1.00 | 0 | 0 | -1 | 0 | 0 | 1 | 0 | 0 | 1 | 2 |
| 1.05 | 0 | 0 | 0 | 0 | -1 | -1 | 0 | 0 | -2 | 1 |
| 1.10 | 0 | 0 | 1 | 0 | 0 | 0 | 0 | 0 | 0 | -2 |
| 1.15 | 0 | 0 | 0 | -2 | 4 | 3 | -1 | 1 | 4 | -8 |
| 0.85 | 1 | 1 | -1 | 1 | 0 | -1 | 2 | 0 | -1 | -3 |
| 0.90 | 2 | 1 | 0 | 0 | 0 | 1 | 1 | 1 | 1 | -1 |
| 0.95 | -1 | -1 | 1 | 0 | 0 | 1 | 0 | 0 | 0 | -1 |
| 1.00 | 0 | 0 | -1 | 0 | 0 | 0 | 0 | -1 | 0 | 1 |
| 1.05 | 0 | 0 | 0 | -1 | -1 | -2 | 0 | -1 | -3 | 1 |
| 1.10 | 0 | -1 | 0 | 0 | 0 | 1 | 1 | 1 | 1 | -1 |
| 1.15 | 0 | 0 | -1 | -2 | 3 | 7 | 2 | 4 | 6 | -7 |

The correct implementation of a Monte Carlo method for valuation of exotic products would require the use of an exact method such as [Broadie et al, 2004]. A regular Monte Carlo implemented with the Euler or even the Milstein method would not correctly work as the Feller condition ($2\kappa\theta > \sigma^2$) is not satisfied. This condition ensures that the variance process cannot reach zero. When the variance process reaches zero, an absorbing boundary condition is imposed. The discretization of the Monte Carlo cannot properly mimic this continuous absorbing condition and options are considerably overpriced. This bias increases with maturity and is not significantly reduced when the simulation time step shrinks. For the constrained case this bias is around 15% for 10 year maturity and 3% for 1 year maturity with 50 thousand simulations and a time step of 0.1 days. The bias error for the unconstrained case is around 100% for the 10 year maturity and 8% for the 1 year maturity. The error explodes at long maturities because $\sigma$ is a lot higher. From a practical point of view, a constrained calibration that satisfied the Feller condition would be preferred, as conventional Monte Carlo methods would work. However with very skewed market scenarios (as the one analyzed here) this may not always be possible.

## VI. APPLICATION TO FORWARD START OPTIONS

This section applies the methodology of sections II and III for the valuation of forward start vanilla options when the underlying follows Heston's process. This problem has already been addressed by [Lucic, 2003], solving a partial differential equation similar to (56). The results are equivalent to those presented here. However, the approach of this section is straightforward and can be easily implemented and generalized to any model whose evolution can be expressed analitically in terms of a characteristic function of the form (2). Consider the forward start option of equation (34) which fixes the strike at time $t_u$ and expires at time $t_v$ according to Fig. 1.

$$p = P(0,t_v)\mathbf{E}\left((e^{x_v} - Ke^{x_u})^+\right) = P(0,t_v)\mathbf{E}\left(e^{x_u}(e^{x_v-x_u} - K)^+\right) \quad (34)$$

Applying the tower law to equation (34), the premium can be expressed by equation (35), where $\mathbf{x}_u = (x_u, v_u)$.

$$p = P(0,t_v)\mathbf{E}\left(e^{x_u}\mathbf{E}\left[(e^{x_v-x_u} - K)^+ / \mathbf{x}_u\right]\right) = P(0,t_v)E \quad (35)$$

The expectation $E$ can be computed integrating over the state variables $\mathbf{x}_u$ and $\mathbf{x}_v$ at times $t_u$ and $t_v$ respectively as expressed by equation (36), where the integrand has been multiplied and divided by the forward value of the underlying on the forward start date $F_u = \mathbf{E}(e^{x_u})$ to normalize the density function that appears later in (40).

$$E = F_u \int_{\mathbf{R}^2} d\mathbf{x}_u f_{0u}(\mathbf{x}_u/\mathbf{x}_0)\frac{e^{x_u}}{F_u}\int_{\mathbf{R}^2}(e^{x_v-x_u} - K)^+ f_{uv}(\mathbf{x}_v/\mathbf{x}_u)d\mathbf{x}_v \quad (36)$$

The evolution of the increment $\tilde{x}_v = x_v - x_u$ from $t_u$ to $t_v$ depends on the inicial variance $v_u$ but not on the initial value of the underlying $x_u$. Therefore, the distribution of the evolution of $\tilde{x}_v$ starting at zero is the same as the evolution of $x_v$ starting from $x_u$. This fact is expressed by equation (37), where $\tilde{\mathbf{x}}_v = (\tilde{x}_v, v_v)$.

$$f_{uv}(x_v, v_v / x_u, v_u) = f_{uv}(x_v - x_u, v_v / 0, v_u) = f_{uv}(\tilde{\mathbf{x}}_v / 0, v_u) \quad (37)$$

Performing the change of variable $\tilde{x}_v = x_v - x_u$ on (36), yields equation (38).

$$E = F_u \int_{\mathbf{R}^2} d\mathbf{x}_u f_{0u}(\mathbf{x}_u/\mathbf{x}_0)\frac{e^{x_u}}{F_u}\int_{\mathbf{R}^2}(e^{\tilde{x}_v} - K)^+ f_{uv}(\tilde{\mathbf{x}}_v/0, v_u)d\tilde{\mathbf{x}}_v \quad (38)$$

Exchanging the order of integration yields (39), where the inner integral can be interpreted as the density function (40) (the integrand is divided by $F_u$ to make the integral sum 1). Now the expectation $E$ can be expressed in terms of a regular spot vanilla expectation with respect to a different measure: $\tilde{f}$. The problem now is to find the characteristic function of this distribution.

$$E = F_u \int_{\mathbf{R}^2} d\tilde{\mathbf{x}}_v (e^{\tilde{x}_v} - K)^+ \tilde{f}(\tilde{\mathbf{x}}_v/\mathbf{x}_0) = F_u \tilde{\mathbf{E}}((e^{\tilde{x}_v} - K)^+) \quad (39)$$

$$\tilde{f}(\tilde{\mathbf{x}}_v/\mathbf{x}_0) = \int_{\mathbf{R}^2}\frac{e^{x_u}}{F_u}f_{0u}(\mathbf{x}_u/\mathbf{x}_0)f_{uv}(\tilde{\mathbf{x}}_v/0, v_u)d\mathbf{x}_u \quad (40)$$

The definition of the characteristic function of the density $\tilde{f}$ is given by equation (41).

$$\tilde{\varphi}(\mathbf{X}/\mathbf{x}_0) = \int_{\mathbf{R}^2} e^{i\tilde{\mathbf{x}}_v \cdot \mathbf{X}} \tilde{f}(\tilde{\mathbf{x}}_v/\mathbf{x}_0)d\tilde{\mathbf{x}}_v \quad (41)$$

If equation (40) is substituted in (41) and the order of integration is exchanged, expression (42) is obtained. The second integral can be recognized as the definition of the characteristic function $\varphi_{uv}(\mathbf{X}/0, v_u)$ given by equation (21) with $x_u = 0$. Replacing this definition in (42) yields (43). Reordering terms gives (44). Considering that $d\mathbf{x}_u = dx_u dv_u$ allows to identify the integral (44) as the definition of the characteristic function $\varphi_{0u}$ as expressed by equation (45). Replacing this definition by (21) yields the final result (46).

$$\tilde{\varphi}(\mathbf{X}/\mathbf{x}_0) = \int_{\mathbf{R}^2} d\mathbf{x}_u \frac{e^{x_u}}{F_u}f_{0u}(\mathbf{x}_u/\mathbf{x}_0)\int_{\mathbf{R}^2} e^{i\tilde{\mathbf{x}}_v \cdot \mathbf{X}} f_{uv}(\tilde{\mathbf{x}}_v/0, v_u)d\tilde{\mathbf{x}}_v \quad (42)$$

$$= \int_{\mathbf{R}^2} d\mathbf{x}_u \frac{e^{x_u}}{F_u}f_{0u}(\mathbf{x}_u/\mathbf{x}_0)\exp(C_{uv}(\mathbf{X}) + D_{uv,2}(\mathbf{X})v_u) \quad (43)$$

$$= \frac{e^{C_{uv}(\mathbf{X})}}{F_u}\int_{\mathbf{R}^2} d\mathbf{x}_u f_{0u}(\mathbf{x}_u/\mathbf{x}_0)\exp(i(-i)x_u + i(-iD_{uv,2}(\mathbf{X})v_u)) \quad (44)$$

$$= \frac{e^{C_{uv}(\mathbf{X})}}{F_u} \varphi_{0u}\left(-i, -iD_{uv,2}(\mathbf{X})/\mathbf{x}_0\right) \quad (45)$$

$$= e^{-\ln(F_u) + C_{uv}(\mathbf{X}) + C_{0u}(-i,-iD_{uv,2}(\mathbf{X})) + x_0 + D_{0u,2}(-i,-iD_{uv,2}(\mathbf{X}))v_0} \quad (46)$$

Therefore, the final expression (47) of the characteristic function $\tilde{\varphi}$, where $\tilde{C}$ and $\tilde{D}$ are given by (48).

$$\tilde{\varphi}(\mathbf{X}/\mathbf{x}_0) = \exp\left(\tilde{C}(\mathbf{X}) + x_0 + \tilde{D}(\mathbf{X})v_0\right) \quad (47)$$

$$\begin{cases} \tilde{C}(\mathbf{X}) = -\ln(F_u) + C_{uv}(\mathbf{X}) + C_{0u}\left(-i, -iD_{uv,2}(\mathbf{X})\right) \\ \tilde{D}(\mathbf{X}) = D_{0u,2}\left(-i, -iD_{uv,2}(\mathbf{X})\right) \end{cases} \quad (48)$$

The marginal characteristic function of $\tilde{x}_v$ is obtained setting the second component $V$ of $\mathbf{X} = (X,V)$ equal to zero and the price of the forward start option can be easily calculated using the same procedure of section III for vanilla options.

## VII. FORWARD SKEW OF HESTON'S MODEL

This section uses the results from section VI to study the forward skew of both calibrations presented in section V. By forward skew it is understood the volatility surface whose implied volatilities $\sigma_{BS}$ that result from the valuation of the forward start option price (34) using the process (49), match the price of the forward start option when $\tilde{x}_t$ is a Heston process.

$$\tilde{x}_t = \left(\mu - \frac{1}{2}\sigma_{BS}^2\right)dt + \sigma_{BS}dW_t \quad (49)$$

As in section V, the options considered are options on the forward $F_t^P$ with spot equal to $F_0^P$ and adjusted strike prices $KF_0^P/F_0^{T_i}$ for each moneyness and maturity. Prices are not discounted.

Fig. 5 to Fig. 7 present the implied volatility of 3 month, 1 year and 5 year options when the forward start term changes and the constrained calibration of section V is used. The curve at the bottom corresponds to the spot option (the calibrated volatility from table 1). These figures show that the implied volatility and the slope of the skew increase with the forward term. This agrees well with the parameter interpretation of section V in which the market was pricing in increasing volatility, increasing uncertainty for the volatility and increasing skew.

Fig. 8 shows the implied volatility for the 3 month option using the unconstrained calibration. The implied volatility starts to decrease (especially around moneyness greater than 0.95) when the forward start term goes beyond 9 months. If Fig. 8 is compared with Fig. 5 (the same option valued with constrained calibration), the implied volatility for forward terms up to 9 months is very similar. This fact is confirmed by Fig. 10, which shows the price difference in basis points between the 3 month option valued with constrained and the same option valued with unconstrained calibration. For forward start terms up to 9 months, the price differences are below 10 basis points. However, bigger differences appear when the forward start term goes beyond 9 months.

The differences given by both calibrations are explained because the forward start options depend highly on the marginal distribution of the variance on the forward start date. This marginal distribution is not calibrated (only the marginal distribution of the underlying is calibrated). A long maturity option depends a lot less on the strike fixed at start. That is why the 5 year option of Fig. 9 valued with unconstrained parameters behaves much more alike Fig. 7 (the same option valued with constrained calibration).

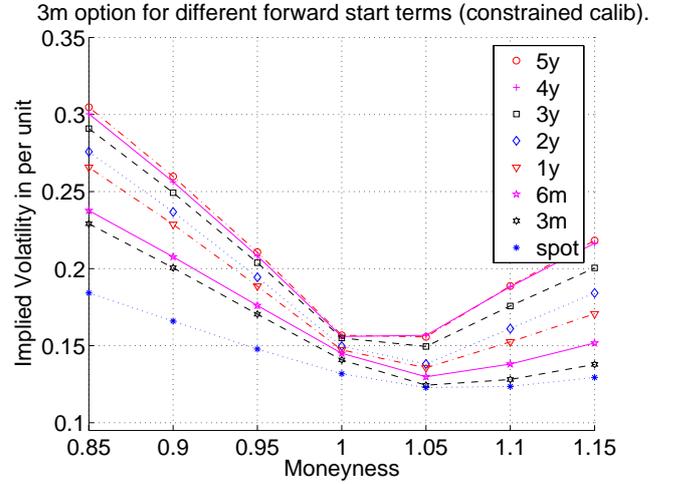

Fig. 5: Implied volatility of a 3 month option for varying forward start terms and using the **constrained** calibration.

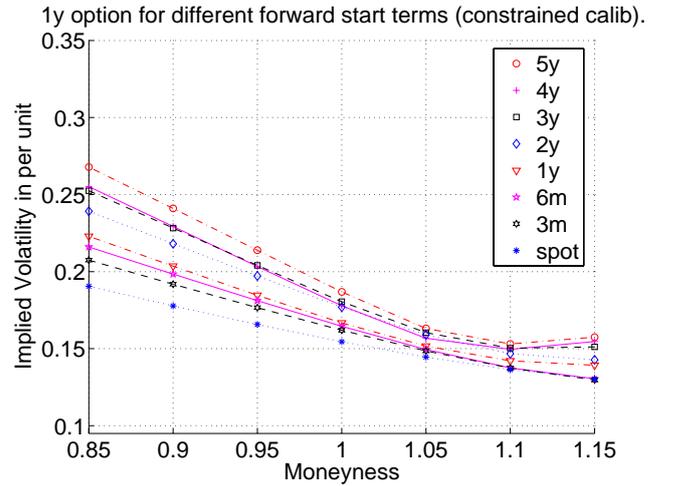

Fig. 6: Implied volatility of a 1 year option for varying forward start terms and using the **constrained** calibration.

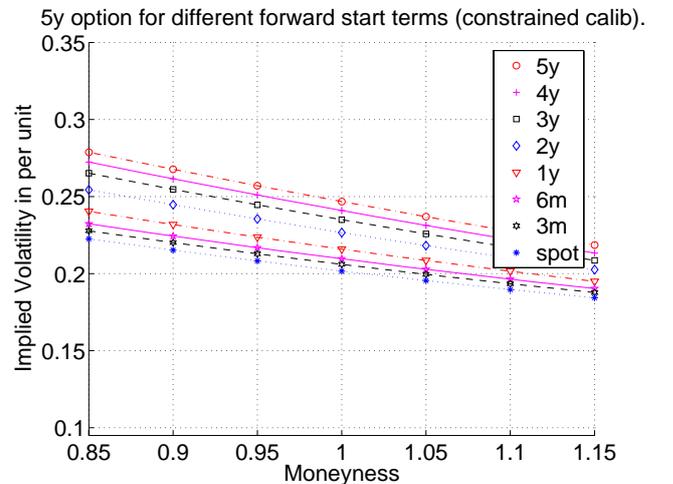

Fig. 7: Implied volatility of a 5 year option for varying forward start terms and using the **constrained** calibration.

The marginal distribution of the variance is a lot more skewed towards lower values for the unconstrained calibration. This fact justifies a lower total variance (the integral of the variance) and therefore lower prices and implied volatilities for the unconstrained calibration. To justify why the unconstrained calibration skews the variance towards lower values, consider the volatility process $\tilde{\sigma}_t$ in (50). This process is obtained applying Ito´s formula to the square root of Heston's variance process $v_t$ in (18).

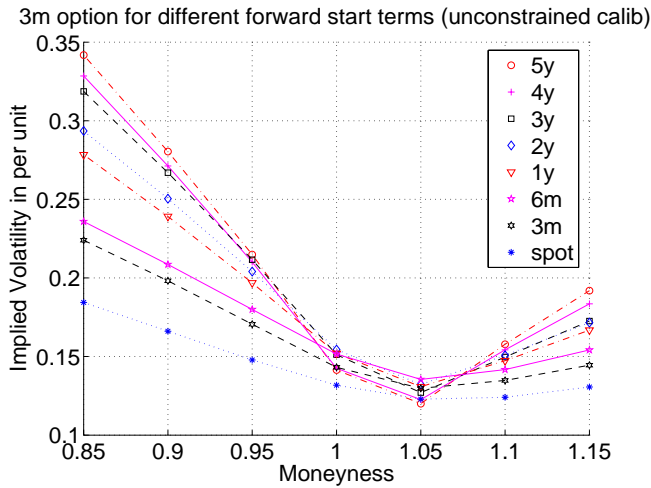

Fig. 8: Implied volatility of a 3 month option for varying forward start terms and using the **unconstrained** calibration.

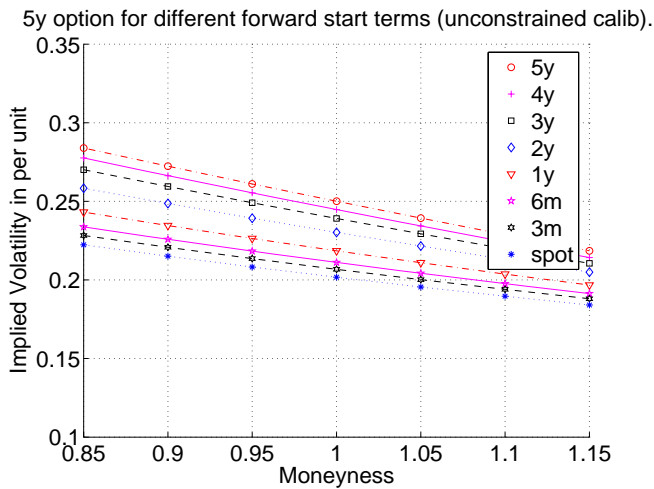

Fig. 9: Implied volatility of a 5 year option for varying forward start terms and using the **unconstrained** calibration.

The unconstrained calibration has greater $\kappa$ and higher $\sigma$ in comparison with the product $\kappa\theta$ for maturities beyond 9 months. This explains why the drift of $\tilde{\sigma}_t$ is considerably more negative for the unconstrained calibration (the drift is negative because the Feller condition is violated and therefore the zero variance point can be reached). See that when $\tilde{\sigma}_t$ equals zero the drift explodes (that´s why this equation cannot be used to integrate the variance process when the Feller condition is not satisfied).

$$d\tilde{\sigma}_t = \left(\frac{4\kappa\theta - \sigma^2}{8\tilde{\sigma}_t} - \frac{\kappa}{2}\tilde{\sigma}_t\right)dt + \frac{1}{2}\sigma\, dY_t \qquad (50)$$

If the forward start option were not so highly dependent on the distribution of the variance, both calibrations would give closer results (as those up to forward start terms of 9 months). However, it seems that the constrained calibration is considerably more reasonable. The results provided in this section show that it is indeed very important to keep in mind that the calibration is carried out only for the marginal distribution of the underlying (that is the only information provided by the market). Therefore, calibration should be carried out so that the forward skew makes sense to traders and practicioners using the model.

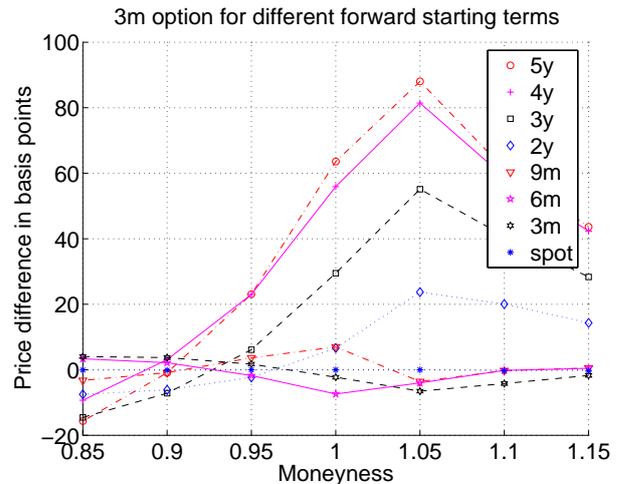

Fig. 10: Constrained minus unconstrained price in basis points of a 3 month option for varying forward start terms.

## VIII. CONCLUSIONS

This paper presents a methodology to introduce piecewise constant time-dependent parameters preserving the analytic tractability for a wide family of models. This family includes hybrids with stochastic interest rates, stochastic volatility and jumps and their respective non-hybrid counterparts.

This methodology is built using transform methods based on analytic expressions of the characteristic function of the distribution of the underlying. The main contribution of the paper is the derivation of the characteristic function of the evolution of the underlying for a time horizon, in terms of the characteristic functions of the horizon sub-periods where the parameters change.

The method is applied to Heston's model to obtain a semi-analytical formula for valuation of vanilla options. A bootstrapping calibration algorithm is proposed and a case study works out the calibration of the volatility surface of the Eurostoxx 50 index.

The method is also applied to obtain a semi-analytical formula for valuation of forward start vanilla options driven by Heston's model. These formulas are used to explore the forward skew of the case study of the Eurostoxx 50 index.

**Acknowledgement**: the author of this paper wants to greatly thank the comments and suggestions of Eduard Giménez. They have improved considerably the organization and the materials presented in this paper. The author also wants to thank Peter Laurence for his helpful comments and Eulogio Cuesta for his notation corrections.

APPENDIX A: DERIVING THE CHARACTERISTIC FUNCTION

Consider the process (51) in terms of $x_t$ and $v_t$ with the same parameter definitions as (18).

$$\begin{cases} dx_t = \left(\mu - \frac{1}{2}v_t\right)dt + \sqrt{v_t}\,dW_t \\ dv_t = \kappa(\theta - v_t)dt + \sigma\sqrt{v_t}\,dY_t \end{cases} \quad d\langle W_t, Y_t \rangle = \rho\, dt \quad (51)$$

The traditional way to calculate the expectation (52) is to integrate the payoff function $g$ using a explicit formula for the density function of the joint probability distribution of $x_T$ and $v_T$ given the initial values $x_t$ and $v_t$. Unfortunately, this density function is not analytic. However, [Heston, 1993] showed that it was possible to calculate the expectation $h$ directly as the solution of a differential equation.

$$h(t, x_t, v_t) = \mathbf{E}\big(g(x_T, v_T)/x_t, v_t\big) \quad (52)$$

The characteristic function of the joint distribution would be given by the function $h$ with payoff function (53).

$$g(X, V, x_T, v_T) = \exp(iXx_T + iVv_T) \quad (53)$$

In the appendix, [Heston, 1993] shows the derivation of the marginal characteristic function using the payoff function $g(X, x_T, v_T) = \exp(iXx_T)$. This appendix provides a more general solution in which the function $g$ can provide payoffs not only of the marginal but also the joint characteristic function. The result presented here can also be found in [Mikhailov et al, 2005] but this paper only mentions that computer-algebra system Maple was used to obtain the result, but no derivation details are provided. Here, the whole derivation procedure is presented.

Consider the function $h$ of equation (54), where $\mathbf{I}_t$ refers to the information set up to time $t$ represented by the values of the stochastic process $x$ and $v$ at time $t$.

$$h(t, x_t, v_t) = \mathbf{E}\big(g(x_T, v_T)/x_t, v_t\big) = \mathbf{E}\big(g/\mathbf{I}_t\big) \quad (54)$$

Considering a time instant $s > t$ and applying the principle of iterated expectations, equation (55) shows that the function $h$ is a martingale.

$$\mathbf{E}\big(h(s, x_s, v_s)/\mathbf{I}_t\big) = \mathbf{E}\big(\mathbf{E}(g/\mathbf{I}_s)/\mathbf{I}_t\big) = \mathbf{E}\big(g/\mathbf{I}_t\big) = h(t, x_t, v_t) \quad (55)$$

Applying Ito's lemma to $h$ and forcing the drift to be zero ($h$ is a martingale) gives the partial differential equation (56). This is indeed a very general result which can be applied to calculate the expectation of functions depending on any process.

$$\frac{\partial h}{\partial t} + \frac{\partial h}{\partial x_t}\left(\mu - \frac{1}{2}v_t\right) + \frac{\partial h}{\partial v_t}\kappa(\theta - v_t) \\ + \frac{1}{2}\frac{\partial^2 h}{\partial x_t^2}v_t + \frac{1}{2}\frac{\partial^2 h}{\partial v_t^2}\sigma^2 v_t + \frac{\partial^2 h}{\partial x_t \partial v_t}\sigma\rho v_t = 0 \quad (56)$$

To determine the solution of equation (56), the final condition (57) at time $T$ must be specified.

$$h(T, x_T, v_T) = g(x_T, v_T) \quad (57)$$

The final payoff function that will be considered has the form (58), where three additional parameters have been introduced: $X$, $C^0$ and $D^0$. As already mentioned in equation (53), if $C^0 = 0$ and $D^0 = iV$, the resulting payoff corresponds to the characteristic function of the joint distribution.

$$g(X, C^0, D^0, x_T, v_T) = \exp(C^0 + D^0 v_T + iXx_T) \quad (58)$$

Equation (59) shows the solution of equation (56) guessed by [Heston, 1993],

$$h(t, x_t, v_T) = \exp(C + Dv_T + iXx_T) \quad (59)$$

where $C$ and $D$ are functions that depend, according to equation (60), on time to maturity $\tau = T - t$, $X$, $C^0$, $D^0$ and all the model parameters in (51), omitted here to simplify notation.

$$C = C(\tau, X, C^0, D^0) \quad D = D(\tau, X, D^0, C^0) \quad (60)$$

Substituting the tentative solution (59) in (56) yields (61), where $A$, $B$ and $M$ are given by (62).

$$-\frac{\partial C}{\partial t} - iX\mu - \kappa\theta D + \left(-\frac{\partial D}{\partial t} + AD^2 + BD + M\right)v_t = 0 \quad (61)$$

$$A = -\frac{1}{2}\sigma^2 \quad B = \kappa - iX\sigma\rho \quad M = \frac{1}{2}X(i + X) \quad (62)$$

As $v_t$ is an independent variable, expression (61) will be zero only if both the term multiplying $v_t$ and the other one are zero independently. On the other hand it is more convenient to use $\tau = T - t$ as parameter rather than $t$. Therefore, the negative of the partial derivatives with respect to $\tau$ will replace the partial derivatives with respect to $t$. This leads to the system of differential equations (63) and (64) (for the purpose of solving it $C$ and $D$ are functions of $\tau$). The terminal condition for this system of equations is given by $C^0 = C(\tau = 0)$ and $D^0 = D(\tau = 0)$ so that $h(T, x_T, v_T)$ becomes the final payoff function (58).

$$\frac{\partial D}{\partial \tau} + AD^2 + BD + M = 0 \tag{63}$$

$$\frac{\partial C}{\partial \tau} - iX\mu - \kappa\theta D = 0 \tag{64}$$

Expression (63) is a Riccati equation that only depends on $D$. This Riccati equation can be turned into the ordinal differential equation (65) through the change of variable $z = (D - D_1)^{-1}$, where $D_1$ is a particular solution.

$$\frac{\partial z}{\partial \tau} + Lz - A \quad \text{with} \quad L = -(2AD_1 + B) \tag{65}$$

The solution of the ordinal equation (65) is given by (66):

$$z = U + W\exp(-L\tau) \quad U = \frac{A}{L} \quad W = \left(\frac{1}{D^0 - D_1} - \frac{A}{L}\right) \tag{66}$$

Undoing the change of variables yields the solution (67).

$$D = \frac{1 + D_1(U + W\exp(-L\tau))}{U + W\exp(-L\tau)} \tag{67}$$

Nothing has been said about the particular solution $D_1$ yet. Indeed, if constant solutions were considered, $D_1$ would be the solution of the second order equation (68) after substituting it in equation (63).

$$AD_1^2 + BD_1 + M = 0 \tag{68}$$

This equation has two solutions: taking the positive square root of the second order equation and substituting in (65) yields $L = -d$, where $d$ is given by equation (70). Taking the negative square root yields $L = d$. The solution used by [Heston, 1993] is $L = -d$. [Albrecher et al, 2007] presents an extensive study proving that both solutions are completely equivalent from a theoretical point of view. However, using $L = -d$ gives plenty of numerical problems (especially for long maturities) as reported in [Kahl et al, 2005], whereas the second solution where $L = d$ avoids them all (see [Albrecher et al, 2007] for a rigorous proof; [Lord et al, 2006] reach the same solution using a different technique under certain parameter restrictions). An intuitive way of realizing that $L = d$ is a better choice is because the exponentials in (67) are decaying. This means that the complex exponential will not oscillate as the maturity increases and the modulus would not explode at long maturities. After simple but tedious algebraic manipulations and choosing $L = d$, equation (67) turns into the final solution (69) where the unknown parameters are given in (70). Please, note that if this expression is compared with that of [Heston, 1993], $d$ will appear with the sign changed (it is not a misprint). In addition, $C^0$ and $D^0$ come out, generalizing the result to allow for more flexible payoffs.

$$D = \frac{\kappa - \rho\sigma Xi + d}{\sigma^2}\left(\frac{g - \tilde{g}e^{-d\tau}}{1 - \tilde{g}e^{-d\tau}}\right) \tag{69}$$

$$\tilde{g} = \frac{\kappa - \rho\sigma Xi - d - D^0\sigma^2}{\kappa - \rho\sigma Xi + d - D^0\sigma^2} \quad g = \frac{\kappa - \rho\sigma Xi - d}{\kappa - \rho\sigma Xi + d} \tag{70}$$

$$d = \sqrt{(\kappa - \rho\sigma Xi)^2 + \sigma^2 X(i + X)}$$

If the solution (69) is substituted in equation (64), the resulting equation is the ordinal differential equation (71), where $\alpha$ and $\beta$ are given in (72).

$$\frac{\partial C}{\partial \tau} - \alpha\frac{g - \tilde{g}e^{-d\tau}}{1 - \tilde{g}e^{-d\tau}} - \beta = 0 \tag{71}$$

$$\alpha = \kappa\theta\frac{\kappa - \rho\sigma Xi + d}{\sigma^2} \quad \beta = iX\mu \tag{72}$$

The solution is given by equation (73), where $K_C$ is a constant that will be calculated to satisfy the terminal condition $C(\tau = 0) = C^0$.

$$C = \int \alpha\frac{g - \tilde{g}e^{-d\tau}}{1 - \tilde{g}e^{-d\tau}}\partial\tau + \beta\tau + K_C = \alpha I + \beta\tau + K_C \tag{73}$$

The indefinite integral $I$ can be calculated doing the change of variable $u = \exp(-d\tau)$ and expanding the result in partial fractions with known integral (a logarithm). Equation (74) shows the final result.

$$I = \frac{-1}{d}\int\left(\frac{(g-1)\tilde{g}}{1 - \tilde{g}u} + \frac{g}{u}\right)du = \frac{g-1}{d}\ln(1 - \tilde{g}e^{-d\tau}) + gd\tau \tag{74}$$

Equation (75) shows the constant from imposing the terminal condition $C(\tau = 0) = C^0$.

$$K_C = C^0 - \alpha\frac{g-1}{d}\ln(1 - \tilde{g}) \tag{75}$$

Replacing (74) and (75) in (73) gives the final result (76).

$$C = i\mu X\tau + \frac{\kappa\theta}{\sigma^2}\left(-2\ln\left(\frac{1 - \tilde{g}e^{-d\tau}}{1 - \tilde{g}}\right) + (\kappa - \rho\sigma Xi - d)\tau\right) + C^0 \tag{76}$$

If this result is compared with that of [Heston, 1993], $C^0$ and $\tilde{g}$ come up (they incorporate the initial conditions $C^0$ and $D^0$). In addition $d$ appears with the sign changed (it is not a misprint) as already discussed for equation (69).

**A. Elices** obtained a PhD in Power Systems engineering at Pontificia Comillas University (Madrid, Spain) and a Masters in Financial Mathematics in the University of Chicago. He is a senior quant team member of the model validation group of the Risk Department at Grupo Santander in Madrid after working in a hedge fund in New York.